# Measurements of spin-orbit interaction in epitaxially grown InAs nanosheets

Furong Fan[1], Yuanjie Chen[1], Dong Pan[2], Jianhua Zhao[2,3], and H. Q. Xu[1,3,*]

[1]*Beijing Key Laboratory of Quantum Devices, Key Laboratory for the Physics and Chemistry of Nanodevices and Department of Electronics, Peking University, Beijing 100871, China*

[2]*State Key Laboratory of Superlattices and Microstructures, Institute of Semiconductors, Chinese Academy of Sciences, P.O. Box 912, Beijing 100083, China*

[3]*Beijing Academy of Quantum Information Sciences, Beijing 100193, China*

*Corresponding author; email: hqxu@pku.edu.cn

(Dated: September 24, 2020)

## ABSTRACT

We report on a low-temperature transport study of a single-gate, planar field-effect device made from a free-standing, wurtzite-crystalline InAs nanosheet. The nanosheet is grown via molecular beam epitaxy and the field-effect device is characterized by gate transfer characteristic measurements and by magnetic field orientation dependent transport measurements. The measurements show that the device exhibits excellent electrical properties and the electron transport in the nanosheet is of the two-dimensional nature. Low-field magnetoconductance measurements are performed for the device at different gate voltages and temperatures, and the characteristic transport lengths, such as phase coherent length, spin-orbit length and mean free path, in the nanosheet are extracted. It is found that the spin-orbit length in the nanosheet is short, on the order of 150 nm, demonstrating the presence of strong spin-orbit interaction in the InAs nanosheet. Our results show that epitaxially grown, free-standing, InAs nanosheets can serve as an emerging semiconductor nanostructure platform for applications in spintronics, spin qubits and planar topological quantum devices.

Low-dimensional III-V narrow bandgap semiconductor nanostructures, such as nanowires,[1,2] nanosheets[3-6] and quantum wells,[7,8] have been extensively investigated in recent years for potential applications in high-speed nanoelectronics,[9,10] infrared optoelectronics,[11,12] spintronics,[13,14] quantum electronics,[15] and quantum computation technology.[16] Among them, InAs and InSb nanowires have been demonstrated to possess excellent electron transport properties and have thus been widely used in realizing field-effect transistors,[17] quantum dots,[18,19] Josephson junctions,[20,21] and topological quantum devices.[22-24] Most intriguing properties in these narrow bandgap semiconductor nanostructures are large Landé g-factors and strong spin-orbit interactions (SOIs).[18,19] It has been predicted theoretically and verified experimentally that a semiconductor nanowire with strong SOI can be used to construct a topological superconducting nanowire, in which Majorana zero modes can be created.[22-28] These Majorana zero modes obey non-Abelian braiding statistics and are of key elements in technology development towards fault-tolerant topological quantum computations.[25,26] However, it is practically difficult to braiding Majorana zero modes in a nanowire device and thus it would become inevitable to consider a planar narrow bandgap semiconductor structure with a strong SOI if a convenient manipulation of Majorana zero modes as required in realizing a topological quantum computation ought to be achieved.[29,30] Recently, free-standing InAs nanosheets have been grown via molecular beam epitaxy (MBE).[5] These two-dimensional (2D) nanostructures are promising for constructing planar topological quantum devices, laterally integrated spin qubits and high-performance spin transistors. The presence of a strong SOI in these nanosheets as required in the above mentioned applications has however not yet been demonstrated experimentally.

In this letter, we report on a low-temperature transport study of a planar field-effect transistor device made from an InAs nanosheet grown via MBE. The device is characterized by gate transfer characteristic measurements and by nanosheet resistance measurements at different orientations of applied magnetic fields. We show that the InAs nanosheet exhibits excellent electron transport properties and the electron transport in the nanosheet is of the 2D nature. We perform low-field magnetoconductance measurements for the nanosheet at different gate voltages and temperatures, and analyze the measurements based on the Hikami-Larkin-Nagaoka (HLN) quantum interference theory.[31] We show that the electron spin-orbit length in the nanosheet is on the order of 150 nm and thus the nanosheet possesses a strong SOI.

The free-standing InAs nanosheets employed in device fabrication for this work are grown on a p-type Si (111) substrate in an MBE system by controlling catalyst alloy segregation.[5] First, a thin silver layer is deposited on the substrate with a clean surface in the MBE chamber and is annealed *in situ* to generate seed particles. The InAs nanosheets are then grown by silver-indium alloy segregation at an indium-rich condition at a temperature of 505 °C for 80 min with the V/III beam equivalent pressure ratio being set at 6.3 (where the indium and arsenic fluxes are $9.3 \times 10^{-7}$ and $5.9 \times 10^{-6}$ mbar, respectively). Figure 1(a) shows a scanning electron microscope (SEM) image of the MBE-grown free-standing InAs nanosheets on the growth substrate. The crystalline structure of the InAs nanosheets is examined by transmission electron microscopy (TEM). Figure 1(b) displays a typical spherical aberration-corrected (Cs-corrected) high-angle annular dark-field scanning TEM (STEM) image of an InAs nanosheet from the same growth sample as those shown in Fig. 1(a) and the inset of Fig. 1(b) shows a corresponding selected-area electron diffraction (SAED) pattern recorded along the $[2\bar{1}\bar{1}0]$ crystallographic direction. It is seen that the InAs nanosheet has a wurtzite (WZ) crystal structure with $\{2\bar{1}\bar{1}0\}$ front and back surfaces.

For device fabrication, the MBE-grown InAs nanosheets are mechanically transferred onto a degenerately n-doped Si substrate, covered by a 200-nm-thick $SiO_2$ layer on top, with predefined markers on the surface. The highly doped Si substrate and the $SiO_2$ layer will serve as the back gate and the gate dielectric in the fabricated devices. A few nanosheets are selected and are located relative to the predefined markers on the substrate by SEM. The contact areas are defined on the selected InAs nanosheets by electron beam lithography. To ensure good contacts between metal and the nanosheets, the contact regions are chemically etched in a deionized water-diluted $(NH_4)_2S_x$ solution. The sample is then immediately loaded into a vacuum chamber where a Ti/Au (5/90 nm) bilayer metal film is deposited by electron beam evaporation. The device fabrication is completed by lift-off process. Figure 1(c) shows a SEM image of a fabricated device measured for this work and the circuit setup for magnetotransport measurements. In this device, as it is shown, four contact electrodes are made. The two inner contact electrodes are about 890 nm apart, which is defined as the conduction channel length ($L$), and the average width ($W$) of the conduction channel is about 650 nm. The thickness of the nanosheet is about 35 nm, as measured by atomic force microscopy (AFM), as shown in Fig. 1(d). The gate transfer characteristics of the device are measured in a two-probe configuration as shown in the inset of Fig. 2(a) by applying a constant source-drain voltage ($V_{sd}$) to the two inner electrodes and detecting the channel (source-drain) current ($I_{sd}$) as a function of the back-gate voltage ($V_{bg}$). However, the magnetotransport measurements are performed in a four-probe configuration using the circuit setup shown in Fig. 1(c) by a standard lock-in technique, in which a 17-Hz AC excitation current ($I$) of 10 nA is supplied between the two outer electrodes and the voltage drop ($V$) between the two inner electrodes is recorded. All the measurements are carried out in a physical property measurement system cryostat equipped with a uniaxial magnet and a rotatable sample holder.

Figure 2(a) shows $I_{sd}$ measured for the device depicted in Fig. 1(c) as a function of $V_{bg}$ (gate transfer characteristics) at $V_{sd} = 1$ mV and different temperatures ($T$). The measurements show that the transport carriers in the InAs nanosheet are n-type and the device exhibits a good gate response at the considered temperatures (2 to 300 K). The off-state current increases with increasing temperature due to an increase in thermally excited carrier density, while the on-state current decreases at high temperatures because of increases in phonon scattering. The conducting channel is open at $V_{bg} = 0$ V and the pinch-off threshold ($V_{th}$) is negative. The InAs nanosheet field-effect transistor thus operates in a depletion mode.

The field-effect electron mobility ($\mu$) in the InAs nanosheet can be extracted from fitting of the measured transfer characteristics to the equation of $I_{sd} = \frac{V_{sd}}{G_s^{-1}+2R_c}$, where $G_s$ is the channel conductance and $2R_c$ represents the contact resistance. The channel conductance is related to the mobility via $G_s = \frac{\mu}{L^2}C_{bg}(V_{bg} - V_{th})$, where $C_{bg}$ is the capacitance of the back gate to the nanosheet channel and $V_{th}$ is the gate threshold for current cutoff. Based on the parallel plate model, $C_{bg}$ can be estimated out from $C_{bg} = \frac{\varepsilon_0 \varepsilon_r S}{d}$ with $\varepsilon_0$ being the permittivity of vacuum, $\varepsilon_r$ the relative permittivity of SiO$_2$, $d$ the thickness of SiO$_2$, and $S$ the area of the nanosheet channel. Using $\varepsilon_0 \sim 8.85 \times 10^{-12}$ F/m, $\varepsilon_r \sim 3.9$, $d \sim 200$ nm, and $S \sim 0.58$ μm$^2$, $C_{bg}$ is estimated to be $9.98 \times 10^{-17}$ F. Figure 2(b) shows the same measured transfer characteristics of the device at $T = 2$ K (red curve) as in Fig. 2(a) and the result of fit (black curve) with $\mu$, $V_{th}$ and $R_c$ as fitting parameters. The fit yields $\mu \sim 2850$ cm$^2$/Vs, $V_{th} \sim -12.2$ V, and $2R_c \sim 370$ Ω at $T = 2$ K. The lower inset of Fig. 2(b) shows the extracted mobility at different temperatures by fitting the measured data shown in Fig. 2(a), where the electron mobility in the InAs nanosheet at room temperature is about 1500 cm$^2$/Vs, larger than that in some other 2D materials such as black phosphorus.[32] The high mobility in the InAs nanosheet demonstrates the potential in electronic applications of the nanosheet. The extracted small contact resistance of $2R_c$~370 Ω and the linear output characteristics of the device [see the upper inset of Fig. 2(b)] at $T = 2$ K mean that good ohmic contacts between the metal and the nanosheet have been obtained.

Figure 3 shows the resistance $R$ of the nanosheet, measured at the four-probe configuration using the circuit setup shown in Fig. 1(c), as a function of the perpendicular component of the magnetic field, $B\cos(\theta)$, at $T = 2$ K and $V_{bg} = 10$ V, for several magnetic field orientations $\theta$, where $\theta$ represents an angle of the applied magnetic field with respect to the normal axis of the nanosheet (see the lower inset of Fig. 3). It is seen that, within small fluctuations, all the measured $R$~$B\cos(\theta)$ curves coincide with each other, indicating that the resistance of the nanosheet depends solely on the perpendicular component of the magnetic field. Thus, the electron transport is dominantly of the 2D nature in the nanosheet. The upper inset of Fig. 3 shows zoom-in plots of the measurements at small perpendicular magnetic field components. Here, a sharp resistance dip is visible near zero field in each measured curve. Thus, the device shows the weak anti-localization (WAL) characteristics at low magnetic fields.[33]

The observation of the WAL characteristics at low magnetic fields implies the presence of SOI in the InAs nanosheet. To determine the strength of the SOI in the nanosheet quantitatively, we have performed low-field magnetotransport measurements of the device with the magnetic field $B$ applied perpendicular to the nanosheet in detail and analyzed the measured magnetoconductance based on the HLN quantum interference theory.[31] Figure 4(a) displays the measured low-field magnetoconductance, $\Delta G = G(B) - G(B = 0)$, at different $V_{bg}$ at $T = 2$ K. Here, it is generally seen that the magnetoconductance shows a peak at zero field, the WAL characteristics, due to SOI in the InAs nanosheet. However, the peak-like structure in the low-field magnetoconductance is gradually weakened as $V_{bg}$ decreases and the magnetoconductance turns to show a broad dip structure, the weak localization (WL) characteristics,[34] when $V_{bg}$ becomes less than −11 V. This gate-tunable WAL-WL crossover in the low-field magnetoconductance, which has been observed in, e.g., InAs nanowires,[35] InSb nanowires,[36] and $Bi_2O_2Se$ nanoplates,[37] is a result of competing for a dominant role in low-field quantum transport between the phase coherence length and the SOI length in the InAs nanosheet.

Various characteristic transport lengths, such as phase coherence length ($L_\varphi$), SOI length ($L_{so}$) and mean free path ($L_e$), in a mesoscopic structure can be extracted by analyzing the low-field magnetoconductance measurements.[31,38] In a 2D diffusion system with strong SOI, the HLN theory shows that the quantum correction to the classical low-field magnetoconductance is given by[31]

$$\Delta G(B) = -\frac{e^2}{\pi h}\left[\frac{1}{2}\Psi\left(\frac{B_\varphi}{B}+\frac{1}{2}\right) + \Psi\left(\frac{B_e}{B}+\frac{1}{2}\right) - \frac{3}{2}\Psi\left(\frac{(4/3)B_{SO}+B_\varphi}{B}+\frac{1}{2}\right) - \frac{1}{2}\ln\left(\frac{B_\varphi}{B}\right) - \ln\left(\frac{B_e}{B}\right) + \frac{3}{2}\ln\left(\frac{(4/3)B_{SO}+B_\varphi}{B}\right)\right], \quad (1)$$

where $\Psi(x)$ is the digamma function, $B$ is an out-of-plane external magnetic field, and $B_i$ ($i = \varphi, so, e$) are the characteristic fields and are related to the characteristic transport lengths via $B_i = \hbar/(4eL_i^2)$.

To extract these characteristic transport lengths in our InAs nanosheet, we fit our low-field magnetoconductance measurements to Eq. (1). Solid lines in Fig. 4(a) are the fitting results. It is seen that good fits are obtained for the measurements at low magnetic fields. Figure 4(b) shows the extracted characteristic lengths, $L_\varphi$, $L_{so}$ and $L_e$, as a function of $V_{bg}$. The extracted $L_\varphi$ exhibits a strong dependence on $V_{bg}$ and decreases from ~460 to ~102 nm as $V_{bg}$ varies from 10 to −12 V. This decrease in $L_\varphi$ with decreasing $V_{bg}$ means a stronger decoherence process, due to reduced Coulomb screening and enhanced electron-electron interaction,[39] at a lower electron density. Furthermore, the extracted $L_\varphi$ is much larger than the thickness of the InAs nanosheet (~35 nm), which is consistent with the fact that the electron transport in the InAs nanosheet is of the 2D nature. The extracted $L_{so}$ shows a weak dependence on $V_{bg}$ and is ~155 nm. This value of $L_{so}$ is comparable to the values extracted in several other III-V narrow bandgap semiconductor nanostructures with a strong SOI, such as InSb nanosheets,[15] InSb nanowires,[36] and InAs nanowires,[35,40,41] where the values of $L_{so}$ are on the order of 150-300 nm, 100-400 nm and 100-200 nm, respectively, and thus indicates the presence of a strong SOI and therefore a strong spin relaxation process in the InAs nanosheet. In the present work, the device is made from a thin WZ InAs nanosheet with $\{2\bar{1}\bar{1}0\}$ surface planes and the transport is most likely along the [0001] crystallographic direction. The spin relaxation process in the nanosheet is dominated by SOI of the Rashba type, originated from the structural inversion asymmetry in the perpendicular direction of the nanosheet.[42,43] Theoretically, the Rashba SOI can be modulated by an external electric field perpendicular to the nanosheet. However, a weak dependence of $L_{so}$ on $V_{bg}$ seen in Fig. 4(b) indicates that the single back gate employed in this device could not efficiently tune the electric field in the nanosheet, although the electron density in the nanosheet has been tuned effectively by the gate. This is because on an open conduction state, the nanosheet would act as a metal layer and thus form a planar capacitor with the Si back gate. As a result, by tuning the voltage applied to the back gate could only efficiently tune the carrier density in the nanosheet but not the electrical field in it.

An important observation in Fig. 4(b) is that with decreasing $V_{bg}$, $L_\varphi$ and $L_{so}$ exhibit a crossover at $V_{bg} \sim -11$ V, where the WAL-WL transition takes place as shown in Fig. 4(a). The extracted $L_e$ is also weakly dependent on $V_{bg}$, consistent with the fact that strength of scattering of the carriers in the nanosheet at this low temperature is predominantly determined by the configuration of impurities, defects and any other imperfections in the device, which is insensitive to a change in $V_{bg}$. Approximately, $L_e$ is ~77 nm, much shorter than the lateral sizes of the InAs nanosheet channel, which is consistent with our assumption made above that the electron transport in the InAs nanosheet is in the diffusion regime.

Figure 4(c) shows the low-field magnetoconductance $\Delta G$ measured for the device at $V_{bg} = 10$ V at different $T$. The measurements show the WAL characteristics in the entire range of the considered temperatures (2 to 20 K). It is also seen that the magnetoconductance peak at zero field is sharp at $T = 2$ K, but with increasing $T$, the sharpness of the peak becomes decreased. Again, these $\Delta G$ data are fitted to Eq. (1). The solid lines in Fig. 4(c) present the fitting results. Figure 4(d) shows the extracted values of $L_\varphi$, $L_{so}$ and $L_e$ as a function of $T$. It is clearly seen that both $L_{so}$ and $L_e$ show weak dependences on $T$, while $L_\varphi$ exhibits a strong dependence on $T$. The observed weak temperature dependence of $L_e$ is as one would naturally expected, because the momentum relaxation of carriers in the nanosheet is dominantly caused by imperfections, such as impurities and defects, in the nanosheet and at the interface between the nanosheet and the gate dielectric, and the configuration of these imperfections in the device structure is insensitive to temperature at this low temperature range. The appearance of the weak temperature dependence of $L_{so}$ is due to the fact that $L_{so}$ is dominantly determined by the electrical field presented in the InAs nanosheet, which is insensitive to a temperature change at this low temperature range. The extracted $L_\varphi$ is up to 460 nm at $T = 2$ K, and is rapidly decreased to 205 nm with increasing $T$ to $T = 20$ K. The dephasing process in the nanosheet at low temperatures is dominated by electron-electron scattering in the Nyquist mechanism with small-energy transfer, i.e., by the fluctuating electromagnetic fields generated by surrounding electrons.[39] The $L_\varphi$ fellows a $T^{-0.49}$ dependence [see the solid line in Fig. 4(d)], which is consistent with our early inference that the electron transport in the InAs nanosheet is of the 2D nature. Strong deviations from this $L_\varphi \sim T^{-0.49}$ dependence are found at $T$ lower than 5 K. These deviations are likely due to that the electron temperature in the nanosheet is higher than the base temperature in this very low temperature range.

In conclusion, a single-gate, field-effect device has been made from an InAs WZ-crystalline nanosheet grown by MBE and the 2D electron transport characteristics of the nanosheet have been studied by magnetotransport measurements. It is found that the spin-orbit length $L_{so}$ in the nanosheet is short, on the order of 150 nm, demonstrating the presence of a strong SOI in the nanosheet. It is also found that $L_{so}$ is insensitive to the gate voltage and temperature. The phase coherence length $L_\varphi$ shows a strong dependence on the gate voltage and a power-law dependence on the temperature, and the dephasing process of electrons in the nanosheet is dominated by the Nyquist type of electron-electron scattering with small-energy transfer. These results show that the InAs nanosheet may serve as an emerging platform for realizing spintronic devices, spin qubits and planar topological quantum devices.

This work is supported by the Ministry of Science and Technology of China through the National Key Research and Development Program of China (Grant Nos. 2017YFA0303304 and 2016YFA0300601), the National Natural Science Foundation of China (Grant Nos. 11874071, 91221202, 91421303, and 61974138), and the Beijing Academy of Quantum Information Sciences (Grant No. Y18G22). D.P. also acknowledges the support from the Youth Innovation Promotion Association of the Chinese Academy of Sciences (Grant No. 2017156).

DATA AVAILABILITY

The data that support the findings of this study are available within the article and from the corresponding author upon reasonable request.

**Figure Captions**

FIG. 1. (a) SEM image (side view) of the MBE-grown free-standing InAs nanosheets on the growth substrate. (b) STEM image of an InAs nanosheet taken from the sample shown in (a), recorded along the $[2\bar{1}\bar{1}0]$ crystallographic direction. The inset shows its corresponding SAED pattern. (c) SEM image (top view) of a fabricated device and circuit setup for magnetotransport measurements. (d) AFM image of the InAs nanosheet in (c).

FIG. 2. (a) Source-drain current $I_{sd}$ measured at a constant source-drain voltage of $V_{sd} = 1$ mV as a function of back-gate voltage $V_{bg}$ (gate transfer characteristics) for the device shown in Fig. 1(c), using the two-probe circuit setup shown in the inset, at zero magnetic field but different temperatures $T$. (b) Gate transfer characteristics of the device at $T = 2$ K (red curve). The black curve represents the result of fit to a standard field-effect model (see the text). The lower inset shows the extracted mobility in the nanosheet as a function of temperature from the measurements shown in (a). The upper inset shows the $I_{sd} - V_{sd}$ output characteristics of the device measured in the two-probe circuit setup shown in the inset of (a).

FIG. 3. Resistance $R$ of the nanosheet measured at $T = 2$ K and $V_{bg} = 10$ V in the four-probe circuit setup as shown in Fig. 1(c) as a function of the perpendicular component of magnetic field $B\cos(\theta)$ at different field orientation angles $\theta$ (see the lower inset for the definition of $\theta$). The upper inset shows zoom-in plots of the measurements at a region of small $B\cos(\theta)$ marked by a black rectangle.

FIG. 4. (a) Low-field magnetoconductance $\Delta G$ measured at $T = 2$ K for the device at various $V_{bg}$. Here, $\Delta G = G(B) - G(B = 0)$ and magnetic field $B$ is applied perpendicular to the nanosheet. The top blue curve displays the measured data at $V_{bg} = -12$ V and all other curves are successively offset by $-0.01e^2/h$ for clarity. The gray solid lines are the fits of the measurements to the HLN theory. (b) Characteristic transport lengths $L_\varphi$, $L_{so}$ and $L_e$ as a function of $V_{bg}$ extracted from the fits of the measured data in (a). (c) Low-field magnetoconductance $\Delta G$ measured for the device at $V_{bg} = 10$ V and different $T$. The top orange curve displays the measured data at $T = 20$ K and all other curves are successively offset by $-0.02e^2/h$ for clarity. Again, the gray solid lines are the fits of the measurements to the HLN theory. (d)

Characteristic transport lengths $L_\varphi$, $L_{so}$ and $L_e$ as a function of $T$ extracted from the measured data in (c). The black solid line is a power-law temperature-dependent fit of $L_\varphi$.

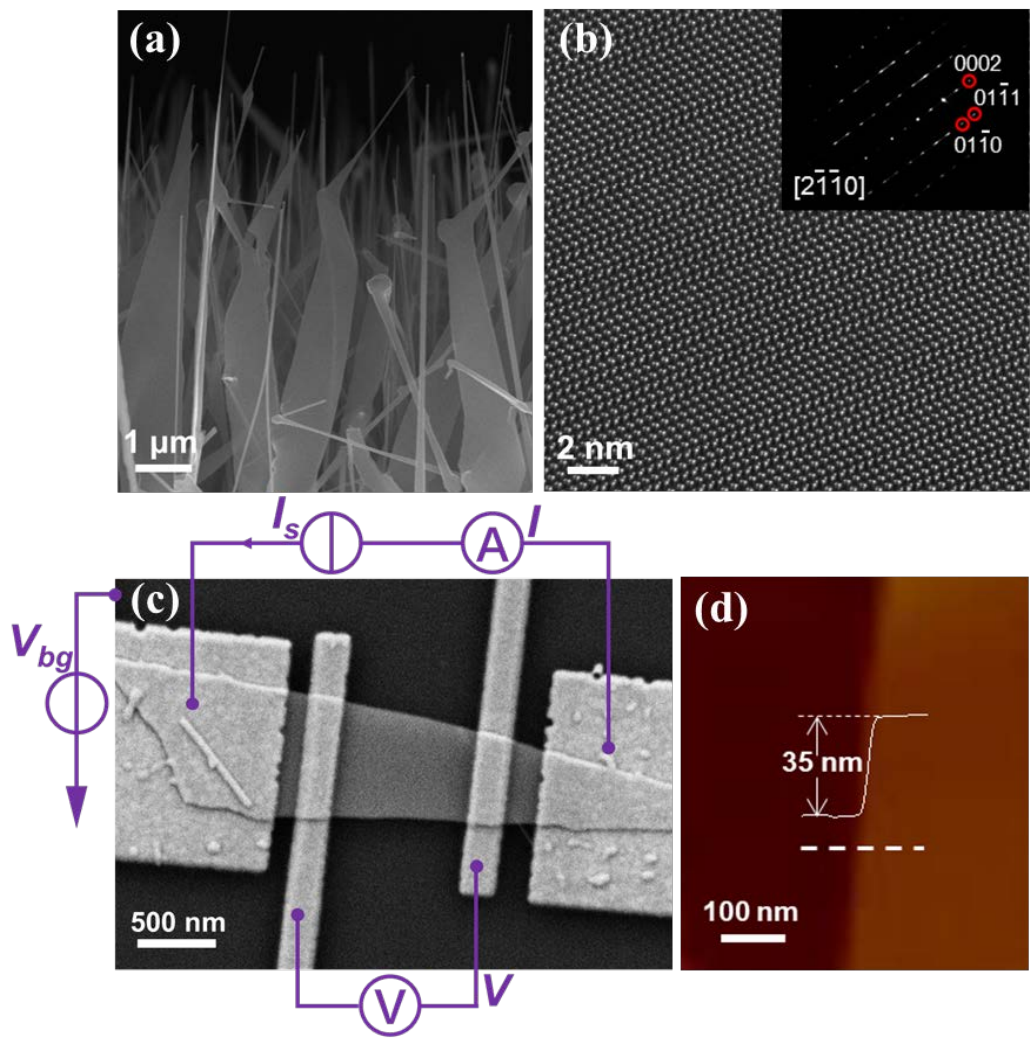

Figure 1, Furong Fan *et al*.

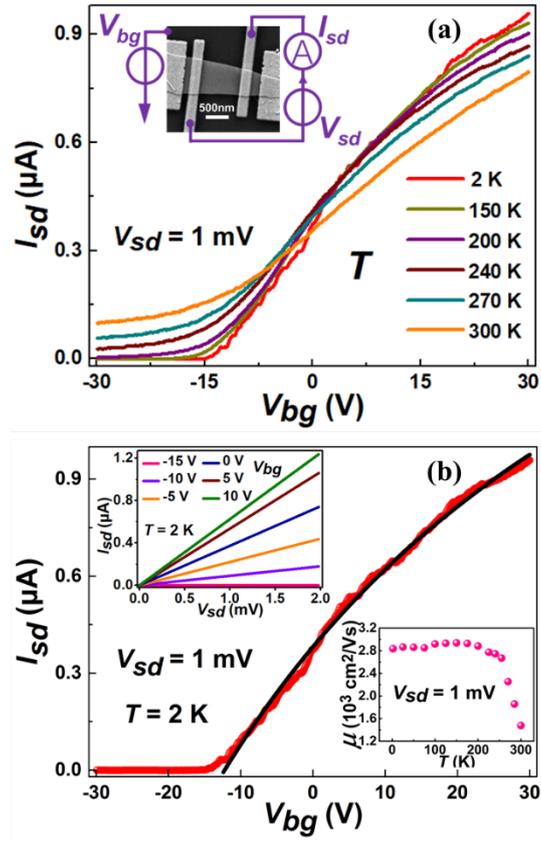

Figure 2, Furong Fan *et al*.

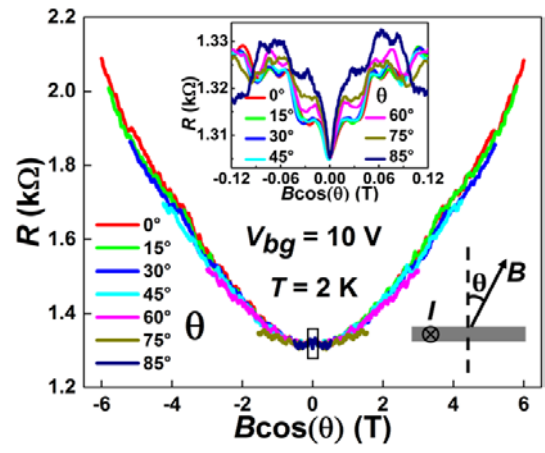

Figure 3, Furong Fan *et al*.

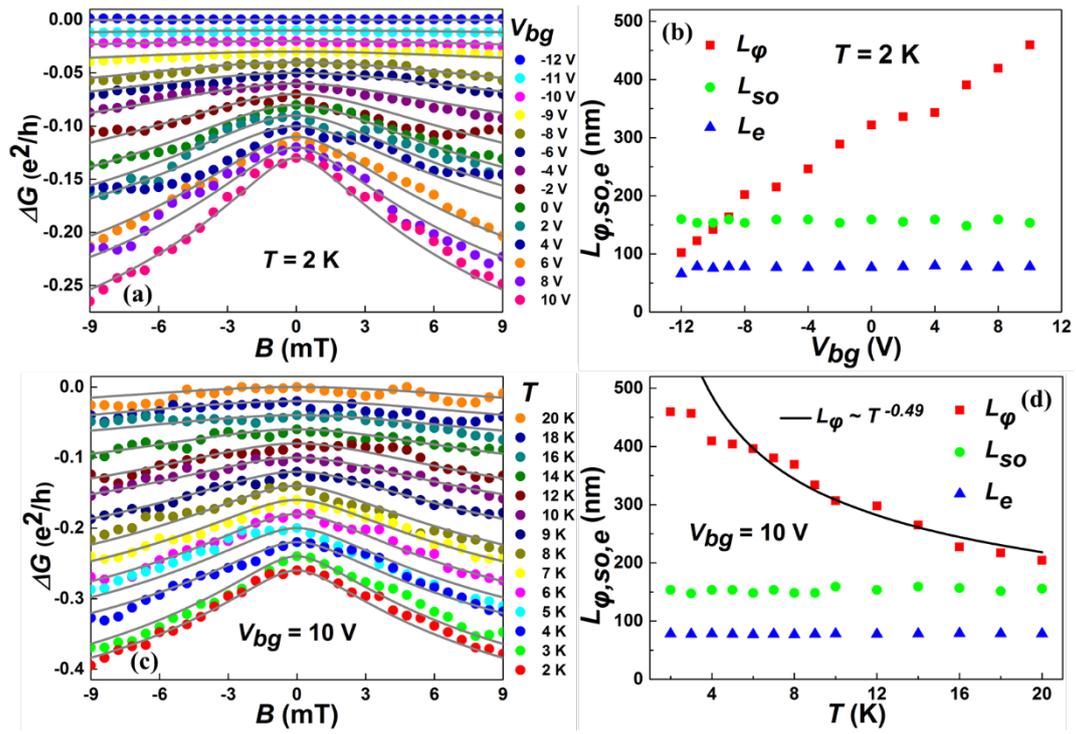

Figure 4, Furong Fan *et al.*